# InGaN/GaN Tunnel Junctions For Hole Injection in GaN Light Emitting Diodes

*Sriram Krishnamoorthy,* [1,a)] *Fatih Akyol*[1]*, and Siddharth Rajan*[1,2,a)]

[1]Department of Electrical & Computer Engineering, The Ohio State University, Columbus, OH 43210

[2]Department of Materials Science and Engineering, The Ohio State University, Columbus, OH 43210

**Abstract:** InGaN/GaN tunnel junction contacts were grown on top of an InGaN/GaN blue (450 nm) light emitting diode wafer using plasma assisted molecular beam epitaxy. The tunnel junction contacts enable low spreading resistance n-GaN top contact layer thereby requiring less top metal contact coverage on the surface. A voltage drop of 5.3 V at 100 mA, forward resistance of $2 \times 10^{-2}$ $\Omega$ cm$^2$ and a higher light output power are measured in tunnel junction LED. A low resistance of $5 \times 10^{-4}$ $\Omega$ cm$^2$ was measured in a MBE grown tunnel junction on GaN PN junction device, indicating that the tunnel junction LED device resistance is limited by the regrowth interface and not by the intrinsic tunneling resistance.

a) Authors to whom correspondence should be addressed.

Electronic mail: rajan@ece.osu.edu, krishnamoorthy.13@osu.edu

This paper describes the incorporation of tunnel junctions on blue light emitting diodes (LED) with minimal electrical loss introduced by the tunnel junction. Tunnel junctions[1] in the gallium nitride material system enable n-type tunneling contacts to p-GaN, multiple active region emitters (multiple wavelength LEDs, multi-junction solar cells) and reverse polarization structures[2] (p-down LEDs). Tunnel junctions also provides a pathway to circumvent efficiency droop in LEDs using a cascaded LED design run at low current density.[3] However, in order to incorporate TJ in commercial devices, low tunneling resistance is a prerequisite. Recently, tunnel junction device designs exploiting polarization in nitrides and embedded rare earth nitride nanoislands resulted in low resistance GaN tunnel junctions with resistivity as low as $10^{-4}$ $\Omega cm^2$.[1,4-7] In case of the polarization[1,4-6,8-11] based approach, a $p^+$-GaN/InGaN/ $n^+$-GaN structure is used where in the depletion field of the $p^+/n^+$ GaN tunnel junction is aligned with that of the polarization field due to the sheet charge at GaN/InGaN interface. Even with a thin InGaN layer (4 nm $In_{0.25}Ga_{0.75}N$), the band edges can be aligned to enable inter-band tunneling.[1] GdN nanoislands embedded in GaN $p^+/n^+$ junction provide intermediate states for tunneling across the depletion region.[7] With reduced tunnel barrier width for each of the two tunneling steps, a resistance of $5 \times 10^{-4}$ – $1 \times 10^{-3}$ $\Omega cm^2$ was reported. Realization of such low resistance GaN-based tunnel junctions has generated interest in tunnel junction enabled devices such as tunnel junction LEDs[12-15], tunnel junction laser diodes[16], and multi-junction solar cells [17,18].

One of the immediate applications of a GaN-based tunnel junction is to replace the high resistance p-GaN layer in commercial LEDs with a n-type top contact. Since the n-type layer has low spreading resistance, the metal electrode coverage can be greatly minimized on the top surface and the light can be extracted from the top surface. Such an approach would have the advantage of avoiding the free carrier absorption losses in ITO current spreading layers in case of top-emitting thin film

LEDs and the complex fabrication processes involved in case of flip-chip LEDs. A tunnel junction draws electrons from the valence band of the p-type material into the n-type material through inter-band tunneling, and holes in the valence band on the p-side are in turn injected into the active region.

Previous reports on tunnel junction LEDs, have demonstrated the advantage in using a low spreading resistance n-type top contact layer. These tunnel junctions were primarily based on degenerately doped PN junctions ($p^+/n^+$ tunnel junctions), [19,20] current spreading by two dimensional electron gas, [21] strained layer super lattices, [22] and transparent conducting ZnO nanoelectrodes.[23] Recently, polarization-based GaN/InGaN/GaN tunnel junctions on LEDs have also been demonstrated using metal organic chemical vapor deposition (MOCVD) technique.[12-15] Although a higher output power was measured in all the above mentioned devices, voltage drop across the tunnel junction was high, and would lead to high losses in LEDs. An ideal tunnel junction device should conduct with minimal bias (voltage drop) across it, while the LED is in forward operation. The on-resistance of the tunnel junction should also be minimal. Tunnel junctions satisfying this criterion, with a resistance of $10^{-3}$ - $10^{-4}$ $\Omega cm^2$ were recently demonstrated by MBE.[1,7] In this work, we demonstrate MBE grown tunnel junctions on commercial LED wafers with the lowest voltage drop is reported for TJ LEDs. Furthermore, previous demonstrations of MBE-grown InGaN/GaN TJs focused on structures along the N-polar orientation.[1,4-6] In this work, polarization engineered InGaN/GaN TJ with lowest tunneling resistance (5 x $10^{-4}$ $\Omega cm^2$) is reported on the technologically more important Ga-polar orientation.

Epitaxial structures of the InGaN/GaN TJ LED and a reference LED without a TJ are shown in Fig 1(b) and 1(c) respectively. The TJ structures were grown by molecular beam epitaxy on an MOCVD grown commercial InGaN/GaN-based blue (450 nm) LED wafer. Standard effusion cells for In, Ga, Si and Mg were used, with

Veeco UNI-Bulb nitrogen plasma source with a RF power of 350 W corresponding to a growth rate of 260 nm/hr. The regrowth interface is generally found to contain oxygen, carbon and silicon impurities, which act as n-type dopants on the surface. Although chemical treatment can reduce the amount of impurities, complete removal has been a challenge, as reported previously for HEMTs.[24] The positive impurity sheet charge concentration at the regrowth interface can lead to a parasitic PN junction in series with the tunnel junction and the LED. With higher sheet charge density of such impurities the resultant undesirable band bending at the regrowth interface can be huge, which is expected to cause additional voltage offset for the TJ LED device. To lower the detrimental effects of this regrowth interface charge induced band-bending, delta doping of Mg was carried out at the bare regrowth interface, keeping the nitrogen shutter closed. This was followed by heavily doped $p^+$-GaN layer using nitrogen shutter pulsing with a duty cycle of 33%. The $p^+$-GaN layer was grown in Ga-rich condition to avoid polarity inversion, and Ga polarity was confirmed through the absence of 3x3 reconstructions[25] in the post-growth RHEED spectrum. Excess droplets were thermally desorbed after the $p^+$-GaN layer to ensure a dry surface before InGaN growth. InGaN was grown in In-rich conditions using a growth model reported elsewhere with a Ga flux less than stoichiometry.[26] By pulsing the nitrogen shutter with a duty cycle of 50%, high doping in $n^+$-GaN ($1.2 \times 10^{20}$ cm$^{-3}$) was achieved. The structure was terminated with an n-type GaN (260 nm, $2.5 \times 10^{19}$ cm$^{-3}$) current spreading layer grown in Ga-rich conditions. To study the effect of the regrowth interface, a reference GaN tunnel junction PN diode shown in figure 1(a) was grown. Growth conditions for the tunnel junction grown on the PN junction was exactly same as that of the regrown tunnel junction shown in figure 1(b). Since the entire sample (TJ PN diode) was grown by MBE, there would be no regrowth interface effects on this device. Figure 2 shows the atomic force microscope (AFM) image of the surface of the InGaN/GaN TJ LED, showing smooth surface morphology with rms roughness in the range of 0.4 - 0.5 nm. Al (20 nm) /Ni (20 nm) /Au (200 nm) top contact metal was e-beam evaporated on $n^+$ GaN using contact lithography. Mesa isolation was done using a $BCl_3$ / $Cl_2$ chemistry to reach n-type GaN bottom contact layer. Al/Ni/Au contacts were deposited on the bottom n-GaN contact layer using e-beam evaporation. Electrical characterization was done using a B1500 Agilent Semiconductor Parameter Analyzer. Electro luminesce and on-wafer light output power measurements were obtained using a calibrated Ocean Optics USB 2000 spectrometer with a coupled fiber optic cable without integrating sphere. A reference LED with no tunnel junction (Fig. 1(c)) was processed using e-beam evaporated semi-transparent Ni (4 nm)/Au (6 nm) contacts on p-GaN and Al/Ni/Au contacts on n-type GaN.

Electroluminescence measurements of the TJ LED and reference sample (pulsed measurement with a duty cycle of 0.1% to avoid thermal heating effects) are shown in Fig. 3. Fig 3 (b) shows a blue shifting of peak emission wavelength from 447.4 nm to 444.9 nm as the current increased from 20 mA to 100 mA in the TJ LED. This is similar to the reference LED where a blue shifting of peak wavelength from 443.8 nm to 442.4 nm was observed when the current increased from 20 mA to 100 mA (Fig. 3 (a)). Since there are no measurable additional peaks in the TJ LED samples, it can be inferred that there is no measurable absorption/ re-emission due to the addition of tunnel junction. It is to be noted that although the band gap of InGaN layer used in the tunnel junction barrier would suggest absorption of blue photons, the generated electron-hole pair would experience a very high electric field in the InGaN layer, as a result of which the generated hole is swept back to the p-GaN layer and electron to the contact. Thus if the internal quantum efficiency of the device is high, which is the case for blue LEDs, the carriers absorbed in the tunnel junction will be re-injected into the active region and absorption losses due to low bandgap InGaN layer in the tunnel junction layer is not expected to be an appreciable loss mechanism.

On-wafer pulsed output power measurements (0.1% duty cycle with) are shown in Fig. 4, showing an increased power output in the case of tunnel junction LEDs, simply because of less electrode coverage on the surface. Optical micrograph shown in Fig. 4(b) shows excellent current spreading in the n-GaN layer. These results are consistent with the previous reports of higher output power in TJ LEDs.[12-14,19-21,23]

The main metric for the TJ integrated on an LED is the electrical loss introduced by the TJ. Current-voltage characteristics of the TJ LED device and the reference LED (350 μm X 350 μm) are shown in figure 5(b). Turn-on voltage, defined as voltage at 100 uA (500 uA) of the InGaN TJ LED is 2.625 V (2.9 V) compared to 2.5 V (2.675 V) in the reference LED. However, after the device turn-on the forward resistance of the TJ LED is comparable to that of the reference. 20 mA (100 mA) current drive, relevant for LED device operation, is achieved at 3.9 V (5.35 V) in InGaN/GaN TJ LED compared to 3.775 V (5.975 V) in the reference sample. The additional turn-on voltage in the TJ LED device is attributed to the barrier for hole injection due to the regrowth interface depletion since no such voltage drops were observed when the entire TJ/PN devices were grown by MBE without a regrowth interface.[1,3,7] Even with the effect of the regrowth interface, the values reported here represent the lowest reported operating voltage for TJ LEDs. Forward resistance of the TJ LED device was extracted to be 2 x $10^{-2}$ $\Omega cm^2$, which is the lowest reported value for any III-nitride tunnel junction LED.

Based on our previous TJ results on PN junctions[1,7], we would have predicted significantly lower operating voltage in the TJ LEDs than that observed here. Depletion due to the positive regrowth interface charge could hence be the reason for the observed additional voltage drop in the device. To verify the effect of the regrowth interface, a GaN TJ PN diode (figure 1(a)) was grown. Identical growth conditions were used in TJ PN diode growth, to achieve identical tunnel junction structures in both the cases. Electrical characteristics of the TJ PN diode device (50 µm x 50 µm) is shown in figure 5(a). The TJ PN diode showed rectification, with a turn-on voltage of ~ 3V, and low series resistance in forward bias. Turn-on voltage indicates no additional voltage drop across the tunnel junction. A total forward resistance of 5 x $10^{-4}$ $\Omega cm^2$ was measured, with the top and bottom contact resistances measured to be 1 x $10^{-6}$ $\Omega cm^2$ and 2 x $10^{-6}$ $\Omega$ $cm^2$ respectively. Hence a tunnel junction resistance of 5 x $10^{-4}$ $\Omega cm^2$ was extracted, which is the total resistance minus the contact resistances. Since the TJ PN diode showed low tunneling resistance as well as no additional voltage drop, the voltage drop measured in the TJ LED structure can be attributed to the regrowth interface depletion.

The regrowth interface is therefore the major challenge that needs to be overcome if the low resistance of tunnel junctions is to be exploited. Growing the entire structure without growth interrupts, such as in an all-MOCVD process would naturally eliminate the regrowth impurities and enable this, while surface treatment strategies previously demonstrated for electronic devices[27] could be adapted for regrown tunnel junctions. Mg out-diffusion and activation of buried p-type layers are also a challenge for integration of tunnel junctions,[12,13] and solutions to these are under active investigation.

In summary, InGaN/GaN tunnel junctions were incorporated in 450 nm LEDs to realize a p-contact free LED design with low spreading resistance n-GaN layer as the top contact layer. The reported operating voltage of 5.3 V at 20 mA is the lowest for GaN LEDs with TJ based p-contacts. Higher output power was achieved in TJ LEDs on account of the low metal footprint on the top surface and low spreading resistance of the top n-type GaN contact layer. MBE grown TJ PN diode, without any regrowth interface, along the Ga-polar orientation, resulted in lowest reported Ga-polar TJ resistance of $5 \times 10^{-4}$ $\Omega$ cm$^2$. The elimination of regrowth interface impurities through an all MOCVD process, or surface treatment procedures will allow LEDs to harness the full potential of polarization engineered tunnel junctions for n-type tunneling contacts and multi-active region structures.

**Acknowledgement**:

The authors would like to acknowledge Prof. James Speck at the Solid State Lighting Center for providing the MOCVD LED wafers used for tunnel junction regrowth in this study. We would like to acknowledge funding from Office of Naval Research under the DATE MURI program (Program manager: Paul Maki), and the National Science Foundation (DMR-1106177).

**Figures Captions**

**Figure 1**: (Color online) (a) Epitaxial stack of tunnel junction (TJ) PN diode structure grown completely by MBE. TJ acts as a tunneling contact to p-GaN. (b) Epitaxial stack of InGaN/GaN TJ regrown on commercial blue LED. Regrowth interface is marked with a red dotted line. (c) A reference stack without the tunnel junction is also shown. Higher the regrowth interface charge sheet density, higher is the barrier for holes injected from the tunnel junction into the LED.

**Figure 2**: (Color online) AFM image of GaN/InGaN TJ LED showing smooth morphology indicating step flow growth with low rms roughness (0.4-0.5 nm).

**Figure 3**: (Color online) Electroluminescence characteristics of (a) reference LED, and (b) InGaN TJ LED.

**Figure 4**: (Color online) (a) On-wafer output power of TJ LED and the reference LED, showing higher output from TJ LED due to less metal footprint on the top contact surface (b) Optical micrograph showing uniform light emission at high current density (100 A/cm$^2$) due to low spreading resistance n-type top layer enabled by the tunnel junction.

**Figure 5**: (Color online) (a) Linear current voltage characteristics of Ga-polar GaN TJ PN diode showing the lowest reported tunnel junction resistance of 5 x 10$^{-4}$ Ω cm$^2$, with negligible additional turn-on voltage. **Inset:** Log J- V characteristics of TJ PN diode. (b) Linear I-V characteristics of the tunnel junction LED and the reference LED (350 μm x 350 μm). On resistance of the TJ LED was measured to be 2 x 10$^{-2}$ Ω cm$^2$. The on resistances are comparable, with additional voltage dropped across the tunnel junction, which is attributed to the depletion region due to the regrowth interface. TJ LED shows lowest reported voltage drop of 5.3 V at 100 mA current drive. **Inset:** Log I- V characteristics of TJ LED.

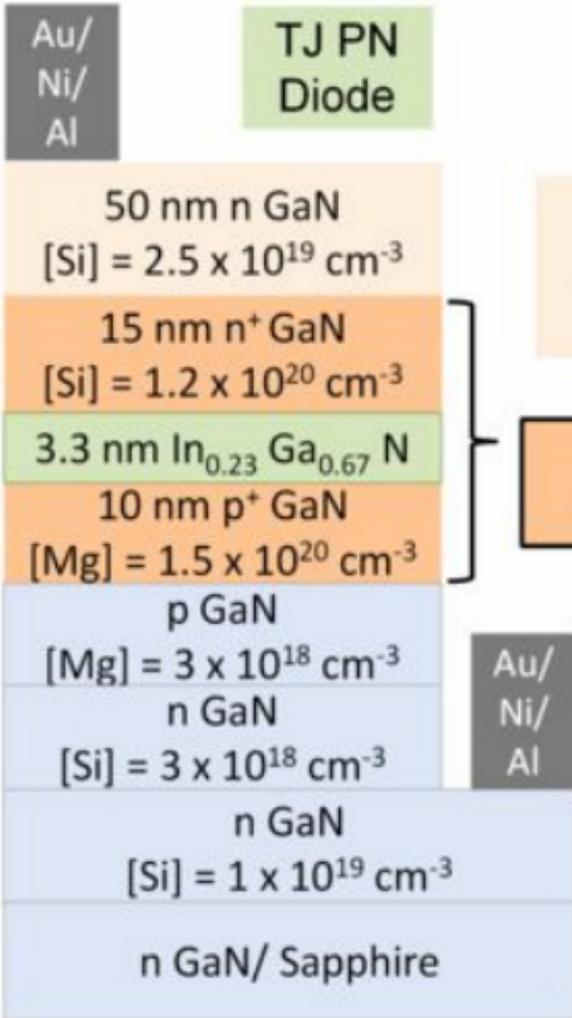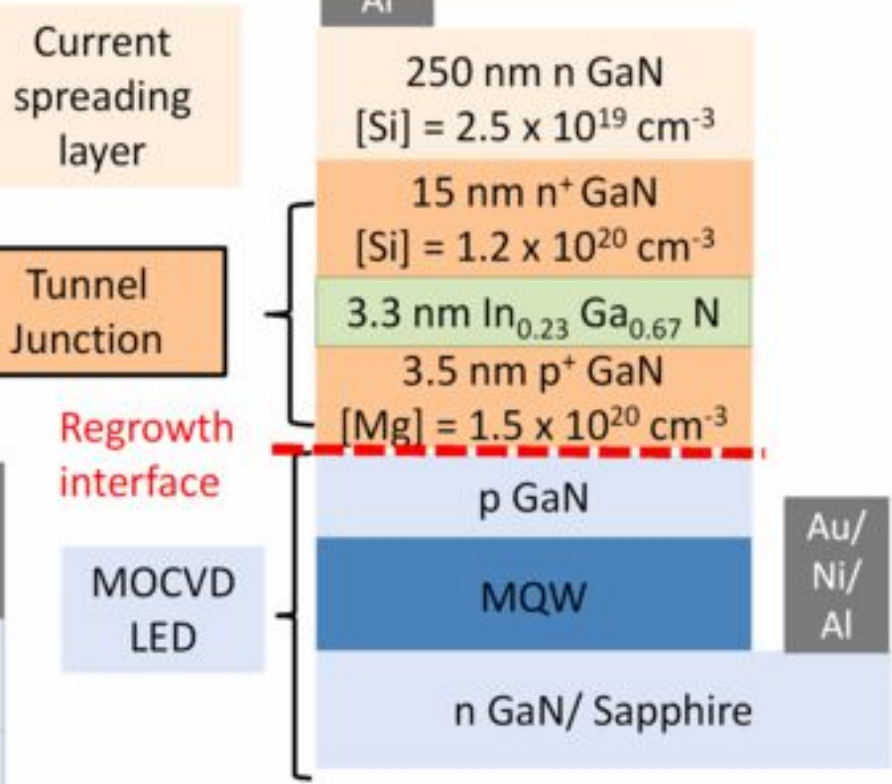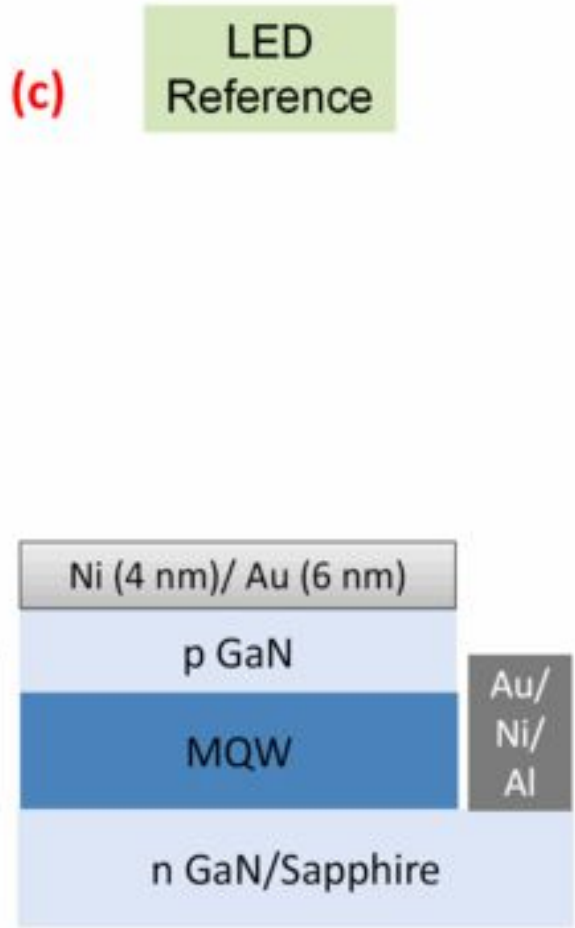

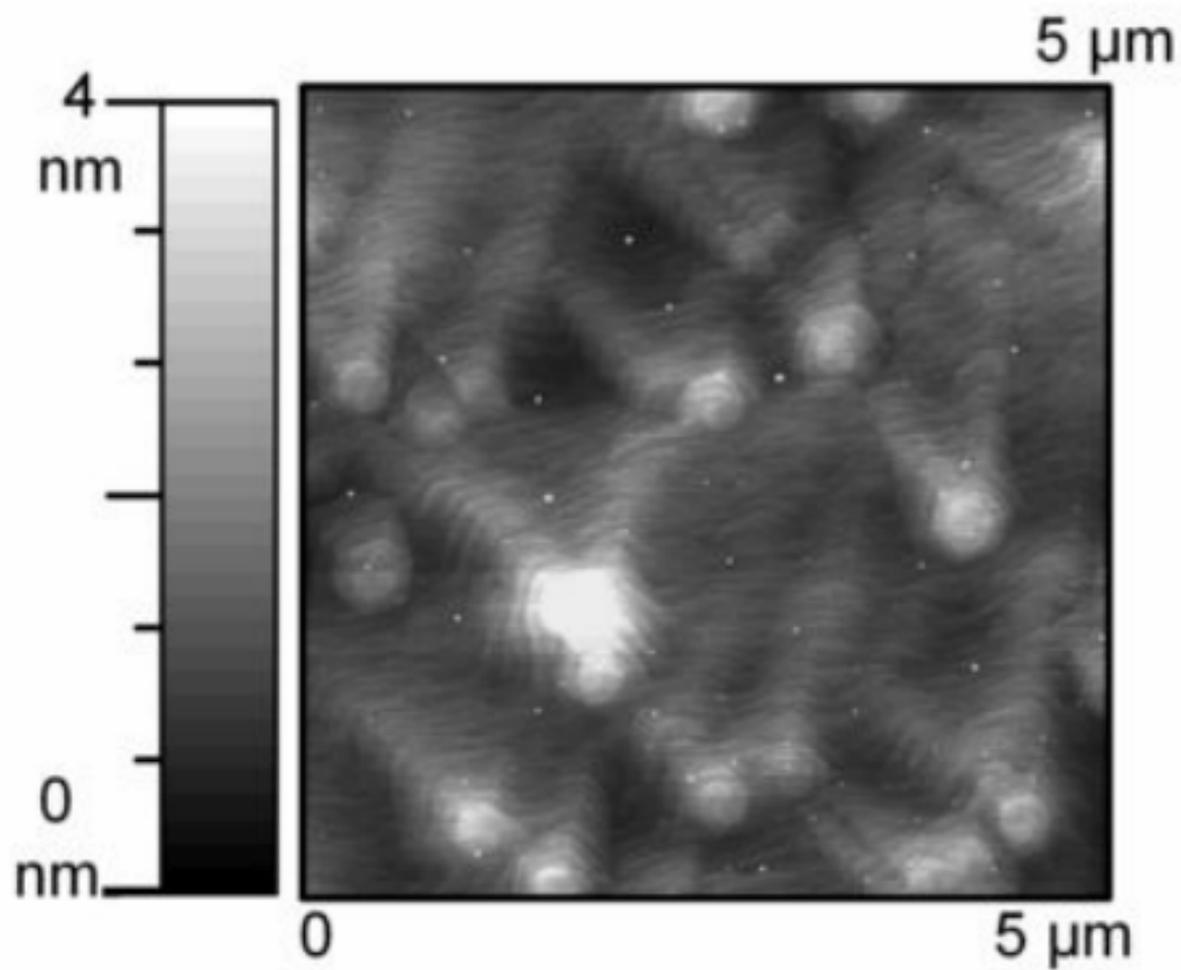

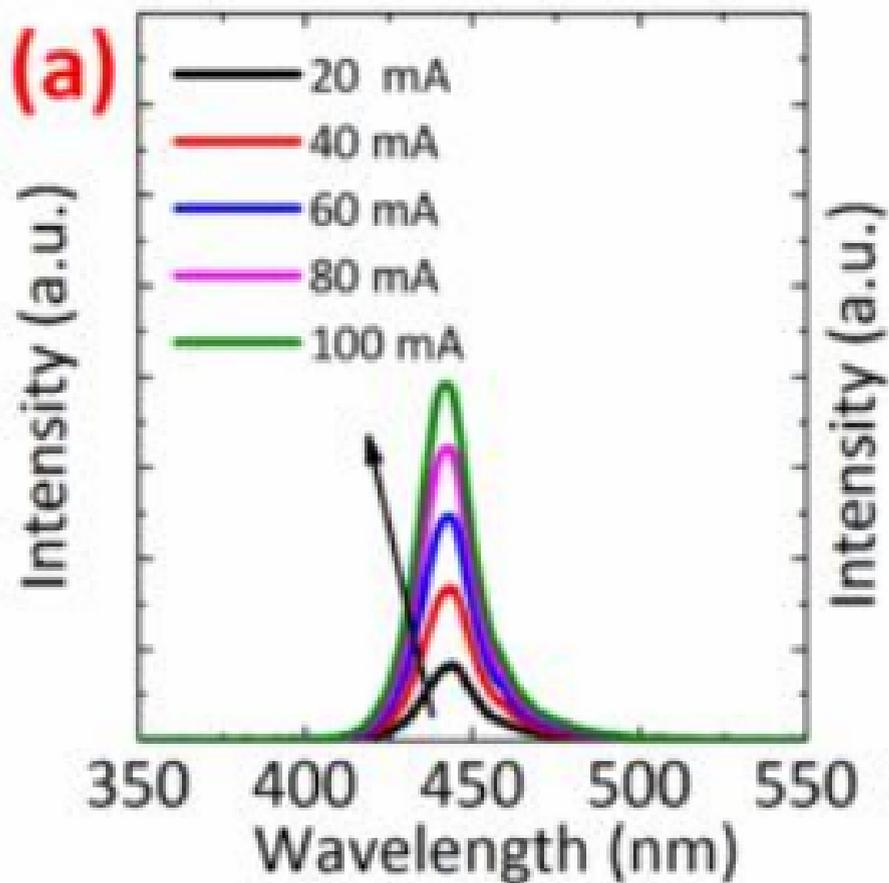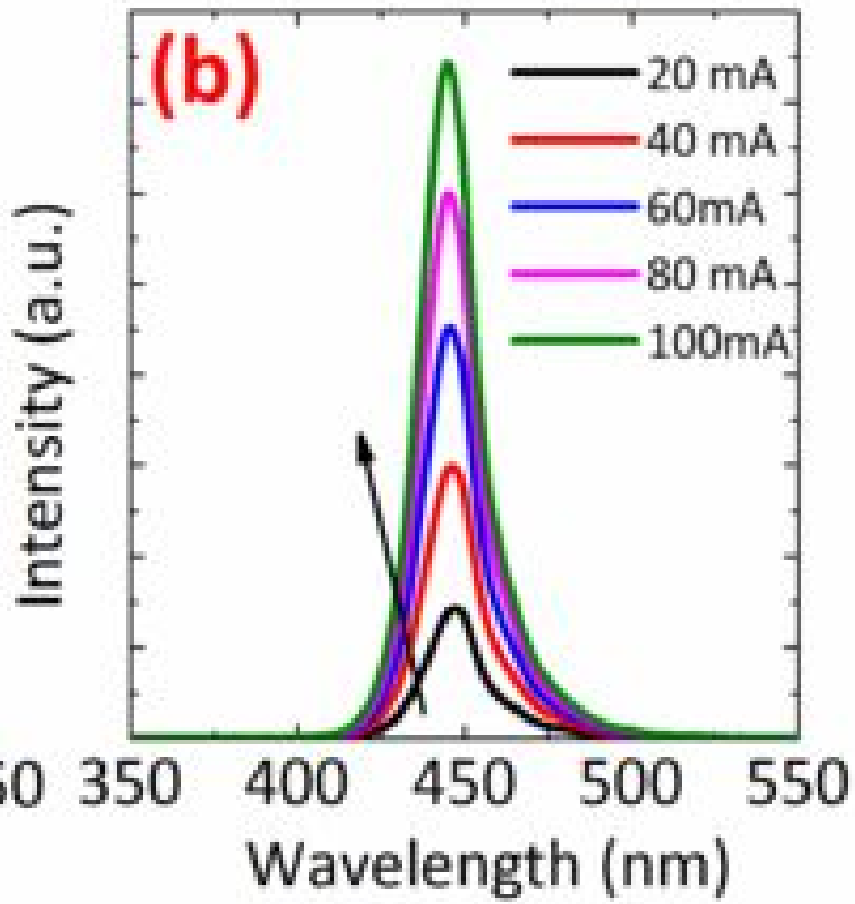

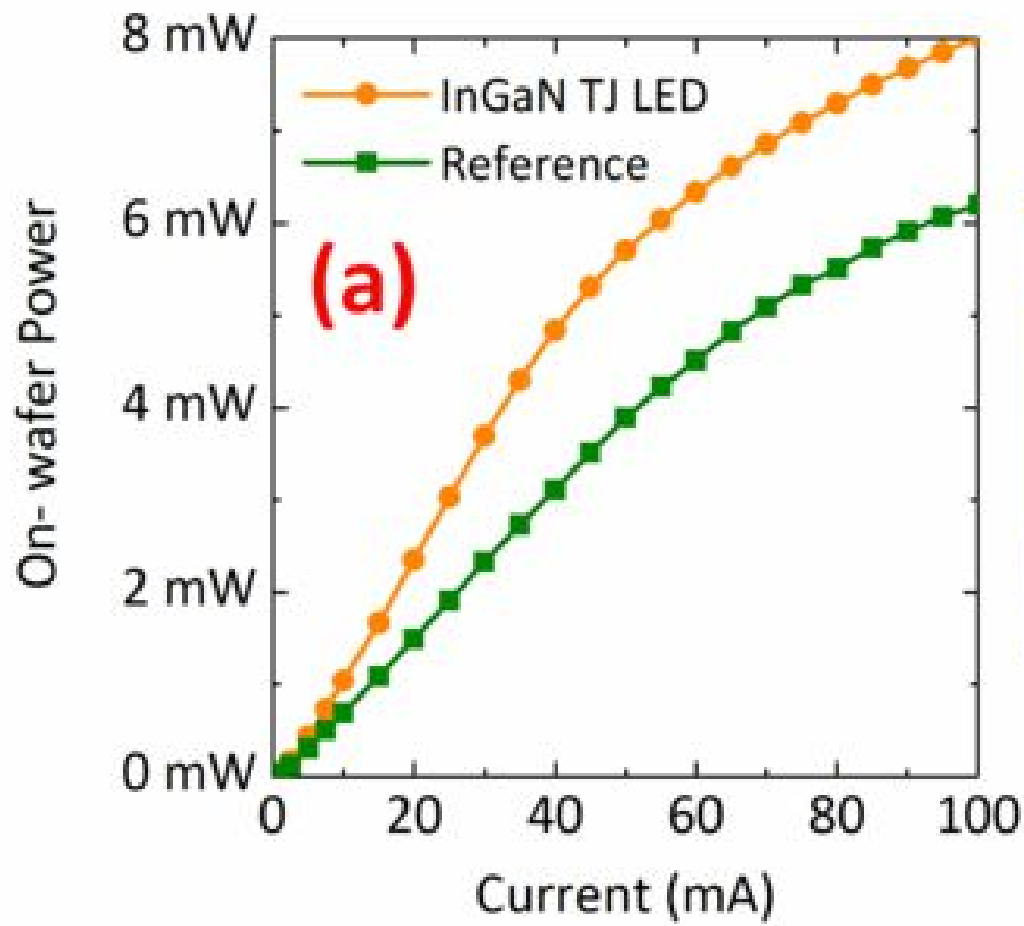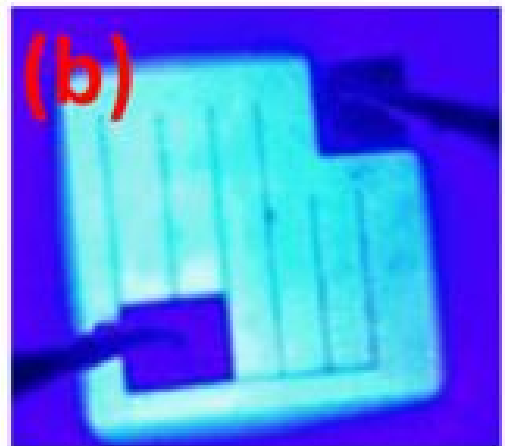

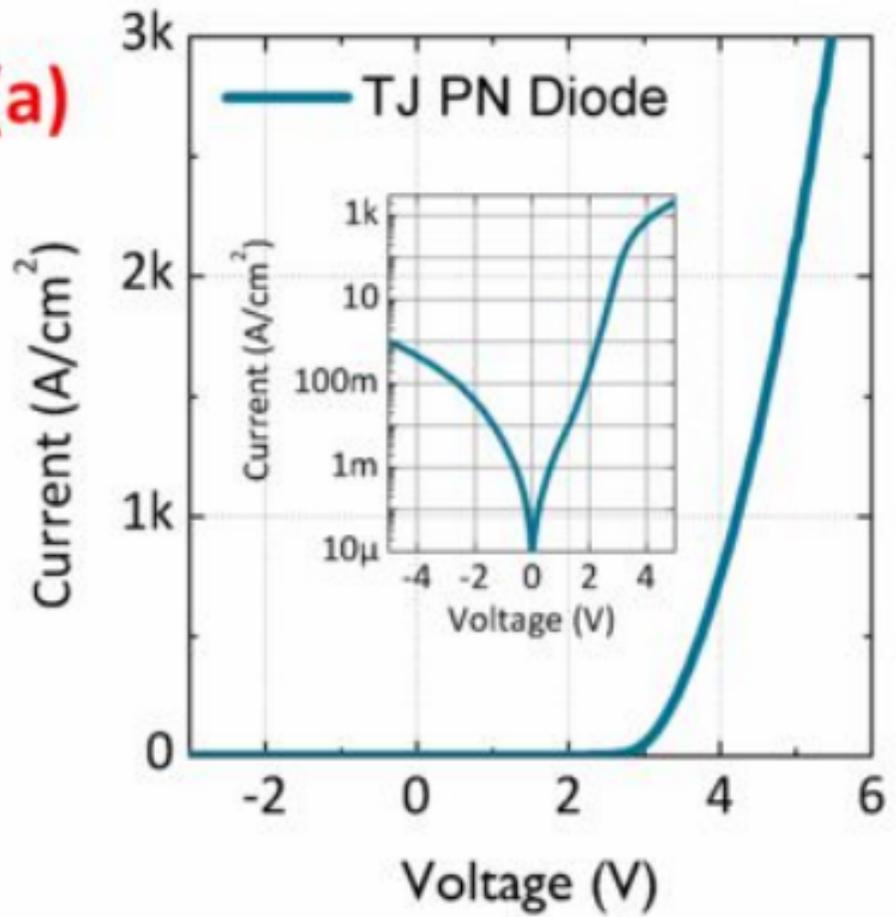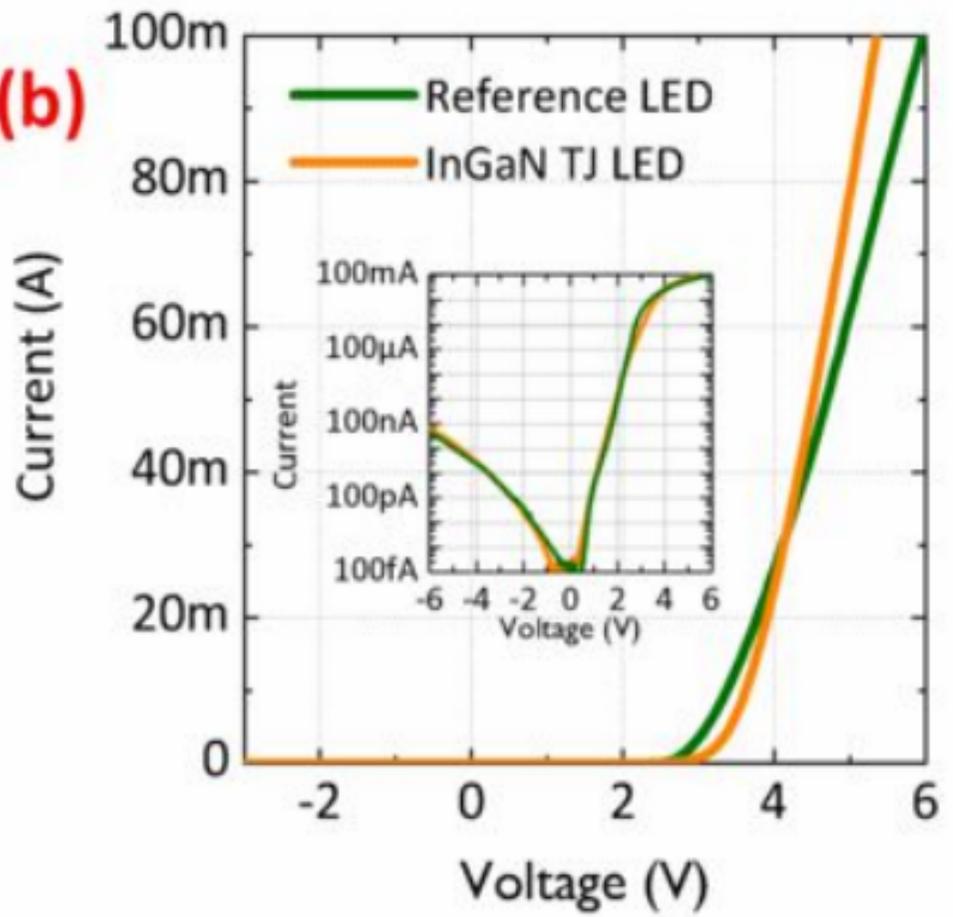